\documentclass[lettersize,journal]{IEEEtran}
\usepackage{amsmath,amsfonts}
\usepackage{amssymb}
\usepackage{array}
\usepackage{threeparttable}
\usepackage{stmaryrd}
\usepackage{hyperref}
\usepackage{textcomp}
\usepackage{adjustbox}
\usepackage{stfloats}
\usepackage{url}
\usepackage{verbatim}
\usepackage{algorithmicx}
\usepackage[ruled,vlined, linesnumbered]{algorithm2e}
\makeatletter
\newcommand{\removelatexerror}{\let\@latex@error\@gobble}

\usepackage{caption}
\usepackage{cite}
\usepackage{xspace}
\usepackage{multirow}
\usepackage{xcolor,soul,framed} %,caption
\colorlet{shadecolor}{yellow}
\usepackage{graphicx}
\usepackage{subcaption} % For subfigures
\usepackage{booktabs} % For formal tables
\usepackage{siunitx} % For unit alignment
\usepackage[utf8]{inputenc}
\DeclareUnicodeCharacter{FF0C}{,}
\begin{document}
\title{Communication-Efficient MARL for Platoon Stability and Energy-efficiency Co-optimization in Cooperative Adaptive Cruise Control of CAVs

}
\author{Min Hua, Dong Chen, Kun Jiang, Fanggang Zhang, Jinhai Wang, Bo Wang, Quan Zhou, and Hongming Xu
% \thanks{

% } 
%Commerce under Grant BS123456.'' }
\thanks{The work is supported in part by the Fundamental Research Funds for the Central Universities (22120240223), by the EPSRC (EP/J00930X/1), and by Innovate UK (102253). }
\thanks{Min Hua, Fanggang Zhang, and Quan Zhou, and Hongming Xu are with the School of Engineering, University of Birmingham, Birmingham, UK. Corresponding author: Hongming Xu (h.m.xu@bham.ac.uk).}
\thanks{Dong Chen is with the Environmental Institute \& Link Lab \& Computer Science, University of Virginia, VA, USA}
\thanks{Kun Jiang is also with with the School of Automation, Southeast University, Nanjing, China}
\thanks{Jinhai Wang is with Hubei Key Laboratory of Advanced Technology for Automotive Components, Wuhan University of Technology, Wuhan, China.}
\thanks{Bo Wang is with the Automotive New Technology Research Institute, BYD Auto, Shenzhen, China}
\thanks{Quan Zhou is also with the School of Automotive Studies, Tongji University, Shanghai, China}
}

\maketitle

\begin{abstract}
Cooperative adaptive cruise control (CACC) has been recognized as a fundamental function of autonomous driving, in which platoon stability and energy efficiency are outstanding challenges that are difficult to accommodate in real-world operations.
This paper studied the CACC of connected and autonomous vehicles (CAVs) based on the multi-agent reinforcement learning algorithm (MARL) to optimize platoon stability and energy efficiency simultaneously. The optimal use of communication bandwidth is the key to guaranteeing learning performance in real-world driving, and thus this paper proposes a communication-efficient MARL by incorporating the quantified stochastic gradient descent (QSGD) and a binary differential consensus (BDC) method into a fully-decentralized MARL framework. 
% We benchmark the performance of CE-MARL against several non-communicative and communicative MARL algorithms, e.g., IA2C, FPrint, and DIAL, through the evaluation of platoon stability, fuel economy, and driving comfort. Compared with state-of-the-art algorithms, the BDC-
% MARL achieves the largest improvement in energy savings, up to 5.8\%, with an average velocity of 15.26
% m/s and an inter-vehicle spacing (IVS) of 20.76 m.
We benchmarked the performance of our proposed BDC-MARL algorithm against several several non-communicative and communicative MARL algorithms, e.g., IA2C, FPrint, and DIAL, through the evaluation of platoon stability, fuel economy, and driving comfort.  Our results show that BDC-MARL achieved the highest energy savings, improving by up to 5.8\%, with an average velocity of 15.26 m/s and an inter-vehicle spacing of 20.76 m.
In addition, we conducted different information-sharing analyses to assess communication efficacy, along with sensitivity analyses and scalability tests with varying platoon sizes. The practical effectiveness of our approach is further demonstrated using real-world scenarios sourced from open-sourced OpenACC.
\end{abstract}


\section{Introduction}
\label{sec:introduction}
% Background
\IEEEPARstart{C}{ooperative} adaptive cruise control (CACC) is a milestone technology in achieving high-level autonomous driving as it provides a sensor-reach platform with high-level computing resources for advanced AI methods \cite{liu2023systematic}. By enabling coherent communication among connected and autonomous vehicles (CAVs), CACC facilitates the formulation of vehicle platoon and determines the optimal velocity and spacing of the vehicle platoon to smooth traffic flow and reduce overall vehicle energy consumption \cite{boddupalli2022resilient, wang2019cooperative, dai2023bargaining}.
Therefore, CACC is a foundational engineering platform to support the revolution of a smarter, safer, and more sustainable transportation system.

Rule-based methods, optimization-based methods, and learning-based methods were developed to achieve CACC of CAVs. Rule-based methods are simple and easy to implement but have limited adaptability in complex traffic environments. 
Optimization-based CACC methods were developed based on the optimal control theory (e.g., model predictive control (MPC)) to calculate the optimal velocity at each step for CAV platoons\cite{wang2024improving, yang2024research}. Optimization-based methods demonstrated their capability in improving fuel economy and traffic efficiency in predefined driving environments\cite{wang2024improving}. 
However, most of the optimization-based methods were based on a assumption that the lead vehicle in a platoon maintains a constant velocity or a foreknown speed profile, limiting its effectiveness in real-world scenarios where vehicles often accelerate or decelerate randomly. On the other hand, the computational demands of the optimization-based methods are significant, making it hard to be implemented on vehicle onboard control units \cite{8370701}.
Compared to rule-based and optimization-based methods, learning-based methods, developed using deep learning, reinforcement learning, or deep reinforcement learning, offer greater adaptability and self-optimization capabilities. For instance, Lin et al. present a learning-based method using the Deep Deterministic Policy Gradient (DDPG) to improve control performance, reducing the episode cost by up to 5.8\% compared to MPC methods \cite{lin2020comparison}.

Among the learning-based methods, multi-agent reinforcement learning (MARL) demonstrated its advancement over the single-agent reinforcement learning methods and other learning-based methods (e.g., imitation Learning) in many fields including industrial robots and autonomous driving \cite{shuai2024optimal, hua2023multi, chen2024lane}. MARL involves multiple agents learning to make decisions in a shared environment, optimizing their actions based on rewards. These agents can work together or compete, allowing them to solve complex tasks that require coordination and improve system performance \cite{chu2020multiagent}.
 In the development of CACC strategies using MARL, most of existing research focused on the development of learning algorithms and communication protocols. For example,
Liu et al. propose an intermediate solution to improve the traffic efficiency of CAVs by avoiding congestion through coordinated behavior controlled by a deep RL agent with an altruistic reward function \cite{liu2021efficient}.
% Wang et al. developed a cooperative eco-driving system, focusing on how the penetration rate of vehicles affects the energy efficiency of the traffic network and propose a role transition protocol for CACC to switch between a leader and following vehicles in a string \hl{MARL?}\cite{wang2019cooperative}.
Han et al. leveraged information-sharing MARL framework with safe actor-critic algorithm to improve traffic efficiency and safety using shared neighbor data with bounded error and safety guarantees \cite{han2023multi}. However, there has been limited research integrating energy consumption optimization into the CACC framework using MARL. Energy efficiency is a critical aspect of CAV operations, especially with the need for sustainable transportation solutions. 

% why decentrailized
In the realm of CACC, a fully decentralized MARL framework is proposed by practical considerations that align with the operational constraints specific to networked vehicular environments, which minimizes reliance on global observations during execution. In \cite{zhang2021decentralized}, global data collection introduces significant risks, including increased communication delays and higher failure rates, which in turn compromise system robustness. In contrast, Xie et al. propose a decentralized control protocol that relies on local neighbor information, addressing the challenges of unreliable communication and enhancing safety and stability \cite{9933795}. The decentralized framework offers increased robustness by ensuring that the failure of a single vehicle does not compromise the entire network, thereby maintaining continuous system operation and safety \cite{long2023hierarchical}. 
Petrillo et al. utilize an adaptive synchronization-based control algorithm with decentralized mitigation of malicious information, ensuring robustness and resilience against various cyber threats \cite{petrillo2020secure}. Additionally, Du et al. introduce a decentralized model-based policy optimization framework (DMPO) to enhance data efficiency in CAV control \cite{du2022scalable}.
Therefore, this framework is highly advantageous for CACC due to its scalability, robustness, and reduced latency in decision-making. 
From the perspective of communication protocols, there are communicative MARL methods \cite{foerster2016learning, kuutti2020survey} and non-communicative MARL methods \cite{lowe2017multi, foerster2017stabilising}. However, 
for the CACC based on a decentralized MARL framework, inter-agent communication is critical for driving safety, and coordinating platoon stability \cite{zhang2023impacts, chu2020multiagent}. 
In \cite{gronauer2022multi}, the heuristic or direct information-sharing method is employed to lead to inefficient communications.
% There four main categories of inter-agent communicating methods . The first group utilizes non-communicative strategies with centralized critics for stabilizing training, while the second group employs heuristic or direct information-sharing methods, which can lead to inefficient communications \cite{gronauer2022multi}. These MARL algorithms often bombard agents with redundant information, complicating convergence and pattern extraction from shared data \cite{yuan2022multi}. The third group develops learnable communication protocols that integrate communication with action-value estimations, and the last group employs communication attentions to manage the communication dynamics among agents based on specific priorities. 
These types are easy to perform delayed information sharing \cite{yan2020cooperative}.

While significant advances have been made in learning algorithms and communication protocols for traffic efficiency and safety, crucial aspects like energy efficiency and communication efficacy have been overlooked. This gap is evident in decentralized MARL frameworks, which are robust and scalable but lack energy-efficient strategies and effective communication protocols. This paper addresses this gap by introducing energy efficiency into CACC problems, enhancing communication protocols, and validating the approach with real-world scenarios.
Therefore, the following contributions and novelities distinguish this paper from the existing works:
\begin{enumerate}
    \item We have introduced energy efficiency into a short-term CACC problem as a decentralized MARL, designing a multi-objective reward function for energy efficiency, stability, and safety. We then benchmark the performance of our proposed MARL algorithm against several other algorithms, including both non-communicative and communicative MARL approaches.

    \item We have integrated the quantized stochastic gradient descent (QSGD) method into our communication protocol and proposed a binary differential consensus (BDC) method to enhance the training process, improving information-sharing efficacy in various CACC scenarios. We also conduct extensive information-sharing analyses to assess the communication efficacy, along with sensitivity analyses and scalability tests with varying platoon sizes.

    % \item We conduct comprehensive experiments, and the results show that the proposed approach consistently outperforms xx  in terms of stability, comfort, and economy efficiency.

    \item The practical effectiveness of our approach is further demonstrated using real-world scenarios sourced from the open-sourced OpenACC. By analyzing these real-world scenarios, we validate our model's applicability and robustness in achieving improved energy efficiency, stability, and safety in real-world CACC operations.
    
    % By analyzing real-data scenarios from OpenACC, we develop and integrate an energy consumption model into our MARL framework, infusing our approach with practical, real-world applicability.
\end{enumerate}

The remainder of the paper is organized as follows: Section \uppercase\expandafter{\romannumeral2} introduces the CACC problem formulation for CAVs. The MARL for the CACC scenario is established in Section \uppercase\expandafter{\romannumeral3}. In Section \uppercase\expandafter{\romannumeral4}, the simulation results and discussion are discussed. Conclusions are drawn in Section \uppercase\expandafter{\romannumeral5}.

\section{Architecture of MARL-based CACC for CAVs and the Optimization Problem}
% \subsection{CACC Problem Formulation}
 The multi-agent reinforcement learning system for CACC of CAVs is illustrated in in Figure \ref{fig1:overall}. The system includes N CAVs that are connected through vehicle-to-vehicle network. We assume that all CAVs in this study are plug-in hybrids equipped with a 70 kW internal combustion engine, two motor/generators (MG1 with a nominal power of 21 kW and MG2 with a nominal power of 135 kW), and a 20 kWh battery \cite{zhang2023cuboid, hua2023energy}.
 Details of the vehicle modelling and vehicle specification can be found in \cite{hua2023energy}.
 % Appendix. 
 However, when considering the dynamics of plug-in hybrids without focusing on energy management between two power sources, the electric motor is utilized for propulsion in a short-term condition without engaging the full capabilities of the hybrid powertrain.
 % , thus temporarily excluding the problem of energy savings within the individual vehicle. 
 
 Within an undirected graph $\mathcal{G}_{t}=(\mathcal{N}, \mathcal{E}_{t})$ at each step $t$, agent $i$ takes into account not only their current state but also the information $\mathcal{E}_{t}$ from their neighbors ($\mathcal{N}_{nei}$ represents the front and rear vehicle).
 % Each CAV is controlled by a dedicated Actor-Critic agent that calculates longitudinal actions at every time step $t$ using its current state 
 % with an undirected graph $\mathcal{G}_{t}=(\mathcal{N}_{nei}, \mathcal{E}_{t})$ at every time step $t$ using its current state 
 and the states of its neighbor vehicles $\mathcal{N}_{nei}$. 
 % $\mathcal{E}_{t}$ and the states of its neighbor vehicles, $\mathcal{N}_{nei}$. 
 To train the MARL system, a new state representation mechanism is proposed by incorporating the long short-term memory (LSTM) network with a fully-connected (FC) neural network for the actor networks and critic networks of all learning agents. To update the $\mathcal{N}$ critic networks effectively, the QSGD method is developed. Details of the system are described as follows:
\begin{figure*}[!ht]
\centerline{\includegraphics[width=0.85\textwidth]{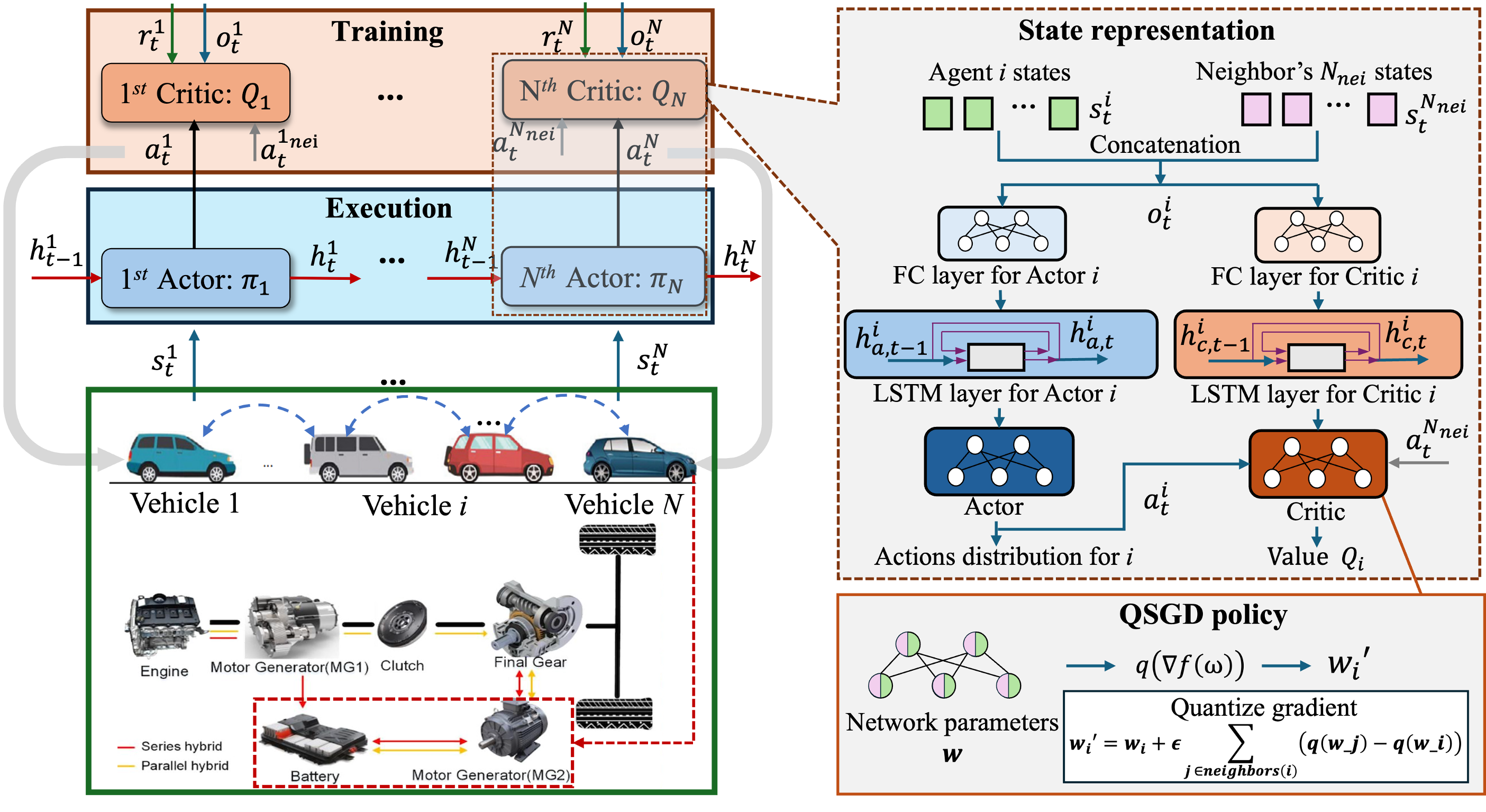}}
\caption{The multi-agent system for CAVs}
\label{fig1:overall}
\vspace{-20pt}
\end{figure*}
\vspace{-10pt}
%%%%%%%%%%%%%% Move Vehicle Modelling to Appendix
\subsection{Decentralized MADRL framework for CACC of CAV}
In this section, we introduce the CACC problem as a model-free multi-agent MDP framework through a decentralized MARL algorithm to maintain a predefined and adjustable inter-vehicle spacing (IVS) $d^{*}$
% $d^{*}=20 \mathrm{~m}$ 
and longitudinal velocity $v^{*}$
% $v^{*}=15\mathrm{~m} / \mathrm{s}$
. Here, a fully decentralized MARL framework provides agents with only partial observations of the environment, concentrating mainly on the vehicles directly ahead and behind. This realistically reflects the practical constraints for each vehicle, which are limited in their ability to sense or communicate with other vehicles. Therefore, a partially observable Markov decision process (POMDP) has been formulated by the tuple $\left(\mathcal{S},\left\{\mathcal{A}_{i}\right\}_{i \in \mathcal{N}}, P,\left\{R_{i}\right\}_{i \in \mathcal{N}},\left\{\mathcal{G}_{t}\right\}_{t \geq 0}\right)$.
More details \cite{chen2024communication} are given as follows:

\subsubsection{\textbf{State Space}}
To achieve an optimal balance between accurately modeling vehicle behavior and avoiding overcomplexity, the state space $\mathcal{S}$ is defined as $\left[v, v_{\text {diff }}, v h, d, u\right]$, specifically, 

\begin{itemize}
    \item The first state is the current normalized vehicle velocity $v=\left(v_{i, t}-v_{i, 0}\right) / v_{i, 0}$ from time $0$ to current time $t$, where $v_{i}$ is longitudinal velocity.
    \item The second state is the clipped vehicle velocity difference with its leading vehicle $v_{\text {diff }}=$ $\operatorname{clip}\left(\left(v_{i-1, t}-v_{i, t}\right) / 5,-2,2\right)$, with $v_{i}$ and the correponding preceding vehicle ($i-1$).
    \item The third state is the IVS-based velocity 
    % defined in Eq. (\ref{eq8}) 
    according to vehicle behavior modeling 
    % in Appendix
    , $v h=$ $\operatorname{clip}\left(\left(v^{\circ}(d_{i})-v_{i, t}\right) / 5,-2,2\right)$, where $v^{\circ}(d_{i})$ represents the behavior of the $i$ th vehicle with optimal velocity model (OVM) \cite{bando1995dynamical} and $d_{i}$ is the IVS of each vehicle $i$.
    The formulation of OVM for the behavior of the $i$th vehicle is defined as follows:
\begin{equation}
\small
\begin{gathered}
       u_{i}=\alpha_{i}\left(v^{\circ}\left(d_{i} ; d^{s}, d^{g}\right)-v_{i}\right)+\beta_{i}\left(v_{i-1}-v_{i}\right) \\
           v^{\circ}(d_{i}) \triangleq \begin{cases}0, & \text { if } d_{i}<d^{s}, \\ \frac{1}{2} v_{\max }\left(1-\cos \left(\pi \frac{d_{i}-d^{s}}{d^{g}-d^{s}}\right)\right), & \text { if } d^{s} \leq d_{i} \leq d^{g}, \\ v_{\max }, & \text { if } d_{i}>h^{g}\end{cases}
\end{gathered}
\label{eq8}
\end{equation}
where $\alpha_{i}$ and $\beta_{i}$ are the IVS gain and relative velocity gain respectively, highlighting how both the IVS and the relative velocity contribute to adjustments in acceleration $u_{i} $ of vehicle $i$ .
% These coefficients are instrumental in mimicking the decision-making process of human drivers, highlighting how both the IVS and the relative velocity contribute to adjustments in acceleration,
Specifically, $d^{s}=5 \mathrm{~m}$ and $d^{g}=35 \mathrm{~m}$ denote the stop IVS and the IVS at full velocity for analyzing the behavior of traffic flow. Additionally, $v^{\circ}$ represents the IVS-based velocity policy. This policy function is continuous and differentiable, which is beneficial for computational models and simulations in the environment of the MARL framework.
    \item The fourth state is the normalized IVS $d=\left(d_{i, t}+\left(v_{i-1, t}-\right.\right.$ $\left.\left.v_{i, t}\right) \Delta t-d^{*}\right) / d^{*}$, where a sampling time $\Delta t$ are given with the discretized longitudinal kinematics for each vehicle $i$.
    \item The last one is the normalized acceleration $u=u_{i, t} / u_{\max }$, where $u_{\max }$ is the maximum acceleration.
\end{itemize}
By normalizing and clipping the values, the model ensures that the state variables remain within practical bounds, enhancing the model's stability and predictability in handling diverse driving scenarios.

\subsubsection{\textbf{Action Space}}
In the defined CACC problem, the action space $\left\{\mathcal{A}_{i}\right\}_{i \in \mathcal{N}}$ is straightforwardly related to the longitudinal control. According to Eq. \ref{eq8}, the OVM control behavior is affected by two critical hyperparameters: IVS gains $\alpha$, relative velocity gain $\beta$.
Thus, these values are chosen from a predefined set comprising four levels: $\{(0,0),(0.5$, $0),(0,0.5),(0.5,0.5)\}$. Subsequently, the longitudinal action is calculated by Eq. (\ref{eq8}).
The selection of four discrete sets of parameters for OVM within a MARL framework is driven by the goal of balancing simplicity, computational efficiency, and the capacity to model a range of driving behaviors for longitudinal vehicle control. These distinct parameter sets simplify the exploration space for the RL, enabling an efficient search through potential actions to optimize driving performance.

%\subsubsection{\textbf{Transition Probabilities}}
%Based on the model-free MARL framework, we do not assume any prior knowledge of this transition probability $P$ while developing our MARL algorithm.
\vspace{-10pt}
\subsection{Co-optimization of platoon stability and energy efficiency}

\subsubsection{\textbf{Reward function}}
\begin{figure}[!ht]
\centerline{\includegraphics[width=0.9\columnwidth]{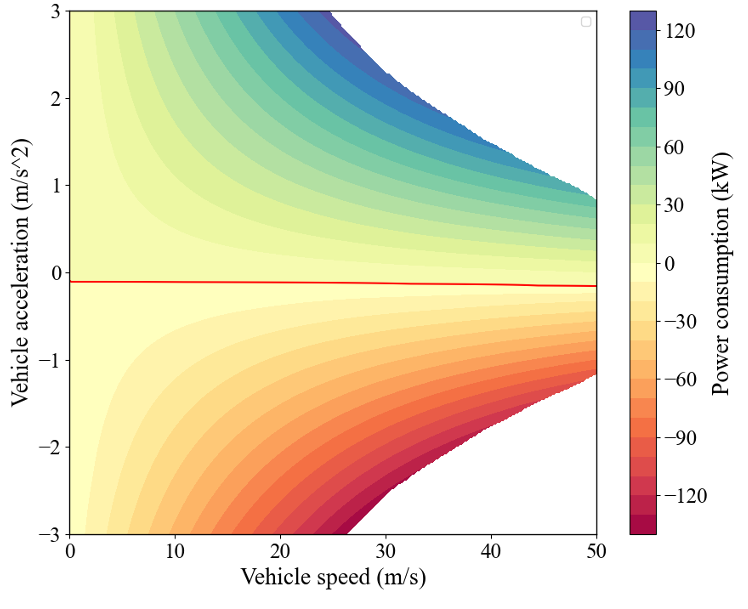}}
\caption{Contour plot of the fitted energy consumption model with respect to vehicle velocity and acceleration.}
\label{fig_power_distribution}
\vspace{-10pt}
\end{figure}

The collective objective is to train the agents toward these desired behaviors. So the multi-objective reward $\left\{R_{i}\right\}_{i \in \mathcal{N}}$ for the $i$ th agent is designed as follows:
\begin{equation}
\begin{gathered}
        R_{i}=  w_{1}\left(d_{i}-d^{*}\right)^{2}+w_{2}\left(v_{i}-v^{*}\right)^{2} 
 +w_{3} u_{i}^{2}\\
 +w_{4}\left(2 d_{s}-d_{i}\right)_{+}^{2}+w_{5} P_{i}
\end{gathered}
\end{equation}
% $$
% \begin{aligned}
% r_{i, t}= & w_{1}\left(d_{i, t}-h^{*}\right)^{2}+w_{2}\left(v_{i, t}-v^{*}\right)^{2} \\
% & +w_{3} u_{i, t}^{2}+w_{4}\left(2 d_{s}-d_{i, t}\right)_{+}^{2}
% \end{aligned}
% $$
where $w_{1}, w_{2}, w_{3}, w_{4}$, and $w_{5}$ are the weighting coefficients. This multi-objective reward design is conducted to achieve a trade-off between performance efficiency, driving comfort, safety, and energy savings in the CACC system. $P_{i}$ is the power consumption of the electric motor of each vehicle, where a polynomial-based, differentiable approximation of an energy consumption model \cite{10359483} is calculated as follows:
% in Appendix.
\begin{equation}
\begin{gathered}
   P_{i} = T_{i} \cdot \omega_{i} \cdot \eta_{i}^{-k} \\
   T_{i}=F_{i}\cdot R 
\end{gathered}
\label{eq}
\end{equation}
where $T_{i}$ represents the motor torque of each vehicle, $\omega_{i}$ is the rotational velocity of the motor for each vehicle, and $\eta_{i}$ is the efficiency of electric-mechanical conversion.
% , detailed in Appendix with Figure \ref{appen2:fig_motor_map}. 
The $k$ indicates the operational mode. $k=1$ denotes the driving model while $k=-1$ denotes regenerative braking mode.  The driving force of each vehicle, $F_{i}$, can be calculated by
\begin{equation}
\begin{gathered}
F_{i} (t)=mgf\cos \alpha +\frac{1}{2} \rho %A_{f}C_{d}v_{i}^{2}+mg\sin \alpha + mu_{i}\\
P_{i}(t)=F_{i}(t)\cdot v_{i} \\
T_{i}(t)=F_{i}(t)\cdot R
\end{gathered}
\label{eq4}
\end{equation}
where $m$ is the vehicle mass; $g$ is the gravity acceleration; $f$ is the rolling resistance coefficient; $\rho$ is the air density; $A_{f}$ is the front area of the vehicle; $C_{d}$ is the aerodynamic drag coefficient; $R$ is the wheel radius; $\alpha$ is the road slope. The critical parameters can be found in \cite{hua2023energy}.
To simplify the calculation and calibration, the vehicle energy consumption, $P_{i}(v_{i}, u_{i})$, is formulated as a function of vehicle velocity $v_{i}$ and acceleration $u_{i}$ as in \cite{guzzella2007vehicle},
\begin{equation}
   P_{i}(v_{i}, u_{i}) = \sum_{k=0}^{4} \sum_{j=0}^{4} p_{kj} \cdot v_{i}^k \cdot u_{i}^j
\end{equation}
where $p_{kj}$ is the polynomial coefficient determined by fitting as shown in Figure \ref{fig_power_distribution}. In this work, each vehicle equipped with the CACC is considered to have identical capabilities and physical characteristics.
The inclusion of the real energy model reveals that the CACC not only regulates vehicle velocity and inter-vehicle spacing (IVS) but also actively manages the powertrain to improve fuel economy and energy efficiency.
% However, when considering the dynamics of the plug-in hybrid without focusing on energy management between two power sources, the electric motor is utilized for propulsion in a short-term condition without engaging the full capabilities of the hybrid powertrain, thus temporarily excluding the problem of energy savings within the individual vehicle. For simplicity, it's assumed that all vehicles possess the same dynamics and capabilities regarding their energy model and their maximum acceleration and deceleration limits.

Further, if the IVS is less than twice the stop IVS $d_{s}$, preventing collisions and ensuring the safety of the vehicle platoon. Here, the "+" in the reward function indicates that a penalty is applied only when the $d_{i}$ falls below twice the $d_{s}$, functioning similarly to a Rectified Linear Unit (ReLU) by activating the penalty under specific conditions; if the IVS $d_{i} \leq 1 \mathrm{~m}$, the substantial penalty of 1000 serves as a strong deterrent against behaviors that could lead to collisions, effectively terminating the training episode to guarantee safety.

\subsubsection{\textbf{Transition Probabilities}}
Based on the model-free MARL, we do not assume any prior knowledge of this transition probability $P$ while developing our MARL algorithm.
\vspace{-10pt}
\section{Communication-efficient MADRL for CACC}

\subsection{Decentralized Multi-agent MDP problem}
The system consists of $N$ agents, indexed by $\mathcal{N}=[N]$, interacting within a shared environment. They communicate via a dynamic network represented as $\mathcal{G}{t}=(\mathcal{N}, \mathcal{E}_{t})$ at each step $t$. The multi-agent Markov decision process (MDP) in this networked environment is defined by the tuple $\left(\mathcal{S},\left\{\mathcal{A}_{i}\right\}_{i \in \mathcal{N}}, P,\left\{R_{i}\right\}_{i \in \mathcal{N}},\left\{\mathcal{G}_{t}\right\}_{t \geq 0}\right)$, where $\mathcal{S}$ is the global state space and $\mathcal{A}_{i}$ is the action space for each agent $i$. The local reward function for each agent $i$ is $R_{i}: \mathcal{S} \times \mathcal{A} \rightarrow \mathbb{R}$, and $P: \mathcal{S} \times \mathcal{A} \times \mathcal{S} \rightarrow[0,1]$ is the state transition probability function.
Each agent $i$ chooses its action $a_{i}$ based on the local policy $\pi^{i}: \mathcal{S} \times \mathcal{A}_{i} \rightarrow[0,1]$, leading to a joint policy $\pi(s, a)=\prod_{i \in \mathcal{N}} \pi^{i}\left(s, a_{i}\right)$. A fully decentralized framework is characterized by the fact that the reward is received locally and the action is executed individually by each agent.The policy for each agent $i$ is parameterized by 
$\pi_{\theta^{i}}^{i}$ with $\theta^{i}$ over $\Theta^{i}$, where $\Theta=\prod_{i=1}^{N} \Theta^{i}$ by packing the parameters together as $\theta=\left[\left(\theta^{1}\right)^{T}, \cdots,\left(\theta^{N}\right)^{T}\right]^{T} \in \Theta$.
Thus, the joint policy then becomes 
$\pi_{\theta}(s, a)=\prod_{i \in \mathcal{N}} \pi_{\theta^{i}}^{i}\left(s, a_{i}\right)$. We assume that the Markov chain $\left\{s_{t}\right\}_{t \geq 0}$ is irreducible and aperiodic under any $\pi_{\theta}$, with the stationary distribution denoted by $d_{\theta}$.
Additionally, the Markov chain of the state-action pair $\left\{\left(s_{t}, a_{t}\right)\right\}_{t \geq 0}$ has a stationary distribution $d_{\theta}(s) \cdot \pi_{\theta}(s, a)$ for any $s \in \mathcal{S}$ and $a \in \mathcal{A}$. Furthermore, the collective objective of the agents is to collaboratively find a policy $\pi_{\theta}$ that maximizes the globally averaged long-term return over the network based solely on local information, formalized as:
\begin{equation}
\begin{aligned}
\max _{\theta} J(\theta) &= \lim _{T} \frac{1}{T} \mathbb{E}\left(\sum_{t=0}^{T-1} \frac{1}{N} \sum_{i \in \mathcal{N}} r_{t+1}^{i}\right) \\
&= \sum_{s \in \mathcal{S}, a \in \mathcal{A}} d_{\theta}(s) \cdot \pi_{\theta}(s, a) \cdot \bar{R}(s, a)
\end{aligned}
\label{eq1}
\end{equation}
where $\bar{R}(s, a)=N^{-1} \cdot \sum_{i \in \mathcal{N}} R^{i}(s, a)$ is the globally averaged reward function. 
The global relative action-value function $Q_{\theta}(s, a)$ and state-value function $V_\theta(s)$ are defined as
$Q_\theta(s, a) = \sum_{t} \mathbb{E}\left[\bar{r}_{t+1} - J(\theta) \mid s_0=s, a_0=a, \pi_\theta\right]$
and
$V_\theta(s) = \sum_{a \in \mathcal{A}} \pi_\theta(s, a) Q_\theta(s, a)$ respectively.
Furthermore, the advantage function can be defined as $A_{\theta}(s, a)=Q_{\theta}(s, a)-V_{\theta}(s)$.  

In this paper,  the typical deployment scenario for CACC involves extensive offline training and testing to address safety and efficiency concerns and avoid online learning. Kuutti et al. emphasize the effectiveness of extensive offline training in reducing deployment risks and enhancing system reliability \cite{kuutti2020survey}. Hence, a decentralized MDP framework ensures that each vehicle, acting as an autonomous control agent, primarily relies on local information and communicates within a limited neighborhood. 
This approach effectively addresses the challenges of partial observability and non-stationarity in a multi-agent environment by allowing each vehicle to enhance responsiveness and adaptability but also ensures that training, although decentralized, happens offline with global information available in batches during the training phase.

\vspace{-10pt}
\subsection{Communication protocol design}
Quantified stochastic gradient descent (QSGD) method is a family of compression schemes with convergence guarantees and practical performance, this method balances communication bandwidth and convergence time by adjusting the number of bits sent per iteration, maintaining convergence even with reduced precision \cite{abdi2020quantized}.
Significantly, the decentralized framework operates without the need for a centralized controller. However, it operates under the assumption that all agents are homogeneous, possessing identical characteristics. While this assumption simplifies the structural complexity of the problem, it fails to capture the inherent diversity among individual agents, especially for CACC. To tackle this issue, we introduce a difference update strategy that promotes a balance between individual learning and the collaborative influence by neighboring agents.
Then, a binary differential consensus (BDC) method is presented, where agents first quantize their weights and then update their weights based on the quantized differential weights between themselves and their neighbors.  Therefore, the update rule is described as:
\begin{equation}
w_i^{\prime}=w_i+\epsilon \sum_{j \in \text { neighbors }(i)}\left(q\left(w_j\right)-q\left(w_i\right)\right)
\end{equation}
where $w_i^{\prime}$ is the updated weight for agent $i$.
 $w_i$ is the current weight of the agent $i$ before the update.
$\epsilon$ is a small positive scalar that scales the update magnitude.
$q\left(w_j\right)$ and $q\left(w_i\right)$ are the quantized weights of neighbor agent $j$ and agent $i$, respectively, mapped to {-1, 0, 1} based on the sign of the weight.
The sum is taken over all neighbors $j$ of agent $i$, as defined by the neighbor mask.

We utilizes quantization to potentially reduce communication overhead while still achieving consensus through differential updates. 
To mitigate the impact of quantization errors, QSGD often incorporates mechanisms like error accumulation (also known as error feedback), which means the quantization error from one iteration (i.e., the difference between the quantized gradient and the actual gradient) is carried over to the next iteration. This error feedback helps compensate for the quantization errors over time, allowing for more accurate updates despite the initial loss of information. The update strategy is given in Algorithm~\ref{algo:quantize}. 

\begin{figure}[!ht]
% \centering
\vspace{-10pt}
\removelatexerror
\scalebox{0.85}{
\begin{algorithm*}[H]
\caption{The gradient descent algorithm with gradient encoding}
\label{algo:quantize}
\SetAlgoLined
\textbf{Data}: Parameter vector $w$ \\
\textbf{Procedure}: Gradient Descent

\For{each iteration $t$}{
Quantize gradient: $q(\nabla f(\boldsymbol{w})) \leftarrow$ Quantize $(\nabla f(\boldsymbol{w})))$; \\
Apply gradient $\boldsymbol{w} \leftarrow \boldsymbol{w}-\epsilon_t q(\nabla f(\boldsymbol{w}))$
} 
\end{algorithm*} } 
\vspace{-10pt}
\end{figure}

In the CACC scenario, by transmitting only three possible values for each gradient component, the system minimizes the amount of data exchanged between vehicles, preserving bandwidth for critical vehicular communications. Since smaller messages mean quicker transmissions, which is crucial for real-time or near-real-time applications like CACC where timely updates can impact system performance and vehicle safety.
With less dependency on the bandwidth, the system can scale more effectively, accommodating more vehicles without proportional increases in communication overhead.
While it introduces some challenges in terms of convergence and accuracy, these can be managed through strategic implementations like error feedback. This makes it a effective technique for distributed systems in high-demand, bandwidth-constrained environments such as CACC scenario.

\section{Experiment setups}
In this section, we evaluate the BDC-MARL framework across various CACC scenarios. Initially, our approach is compared with several leading-edge MARL strategies. Then we explore the efficacy of diverse information-sharing methods. Concurrently, we investigate the effects of different platoon sizes. Finally, the proposed algorithm is applied to scenarios derived from real-world OpenACC data.
\vspace{-10pt}
\subsection{General setups for the learning agents}
The network architecture comprises a single fully-connected input layer for processing state information and an LSTM layer tasked for message extraction, all with hidden layers containing 64 neurons. An orthogonal initializer is utilized to optimize the network parameters during training. 
Training is conducted over $6\times10^{5}$ steps, utilizing a discount factor of $\gamma=0.99$, an actor learning rate of $5.0 \times 10^{-4}$, and a critic learning rate of $2.5 \times 10^{-4}$. 
In the reward function, hyperparameters $w_{1}, w_{2}, w_{3}, w_{4}$, and $w_{5}$ in the reward function are set to $-1.0, -1.0, -0.1$, -5.0, and -10, respectively, with significant emphasis on penalizing scenarios that feature inadequate safe IVS distances. To enhance generalization, each model undergoes training three times using distinct random seeds on a Ubuntu 18.04 server with a NVIDIA RTX 3080. 

%In a simulated traffic environment over a short term period of $T=60s$, we define $\Delta t=0.1s$ as the interaction period between $\mathrm{RL}$ agents and the traffic environment and energy model with pure electric mode. This setup ensures that the environment is simulated for $\Delta t$ seconds following each MDP step. For the experiments, we assume there are a total of 8 $\mathrm{CAVs}$ in the platoon. The impact of different platoon sizes on the performance will be subsequently studied and presented.

%\subsection{CACC scenarios}
%In this paper, we investigate various CACC scenarios to adaptive control the corresponding IVS to a pre-specified value (e.g., $d^{*}=20 \mathrm{~m}$ ) and achieve a target velocity (e.g., $v^{*}=15 \mathrm{~m} / \mathrm{s}$ ) by leveraging real-time $\mathrm{V}2\mathrm{~V}$ communications. The real-world scenario presents a more complex and demanding challenge compared to the simulated environment. This complexity arises from the need for all vehicles to coordinate their velocities and maintain accurate and safe IVS, thereby necessitating more precise control strategies.
\vspace{-10pt}
\subsection{Real-world driving scenario extraction}
The real-world scenarios are from an open-access database involving vehicles with adaptive cruise control (ACC) systems provided by the Joint Research Centre (JRC) of the European Commission (EC) \cite{openacc}. The dataset used in this paper is collected at the ZalaZONE from 10 commercially-available vehicles  with tunable ACC settings. The data was captured using INVENTURE (Race Logic VBOX) equipment, U-blox 9, and the Tracker App, with sensor data from these devices fused and interpolated at a frequency of 10 Hz. 

The real-world driving scenarios are generated for the testing in this paper based on the Dynamic Platform. The Dynamic Platform focuses on the vehicles' response to such perturbations in a controlled environment. This includes a rapid deceleration followed by an acceleration back to the original velocity.
Three groups are selected and extracted respectively from three typical datasets ($\text{dynamic\_part1}$, $\text{dynamic\_part2}$, $\text{dynamic\_part9}$) from timestamps ($316s-376 s, 450s-510 s, 473s-533 s$).
%An energy model was incorporated to evaluate energy consumption resulting from velocity and acceleration adjustments. Data from these experiments were used to assess ACC performance under various driving conditions.

% The evaluation took place on two distinct test tracks: the Dynamic Platform and the Handling Course. These tests assessed the vehicles at low velocitys (8-16 $m/s$) in various platoon configurations under different ACC settings and vehicle sequences at the Dynamic Platform \cite{makridis2021openacc}.
% % , as shown in Fig. \ref{leading_velocity} 
% The Dynamic Platform focuses on the vehicles' response to such perturbations in a controlled environment. This involved a quick deceleration followed by an acceleration back to the original velocity. 
% Three groups are selected and extracted respectively from three typical datasets ($\text{dynamic\_part1}$, $\text{dynamic\_part2}$, $\text{dynamic\_part9}$) from timestamps ($316s-376 s, 450s-510 s, 473s-533 s$).
% Additionally, an energy model was incorporated into the analysis to evaluate the energy consumption resulting from adjustments in velocity and acceleration. Consequently, data collected from experiments using actual vehicles are utilized to assess ACC performance under various driving conditions.

\section{Results and discussion}
In this section, we evaluate the BDC-MARL framework across various CACC scenarios. Initially, our approach is compared with several leading-edge MARL strategies. Then we explore the efficacy of diverse information-sharing methods. Concurrently, we investigate the effects of different platoon sizes. Finally, the proposed algorithm is applied to scenarios derived from real-world OpenACC data.
\vspace{-10pt}
\subsection{Co-optimization Performance with State-of-the-art MARL}
To demonstrate the effectiveness of the proposed method, we compare  with several state-of-the-art MARL algorithms, which include both non-communicative and communicative approaches. Specifically, IA2C \cite{lowe2017multi} employs an independent learning strategy, while FPrint \cite{foerster2017stabilising} integrates neighboring policies into its inputs, both of which are non-communicative methods. In contrast, DIAL \cite{foerster2016learning} and CommNet \cite{sukhbaatar2016learning} implement learnable communication protocols that allow for the inclusion of additional information from neighboring agents, such as states or policy details, which necessitates a higher communication banwidth. All the algorithms compared utilize the same deep neural network architecture.

\begin{figure}
\centerline{\includegraphics[width=1.0\columnwidth]{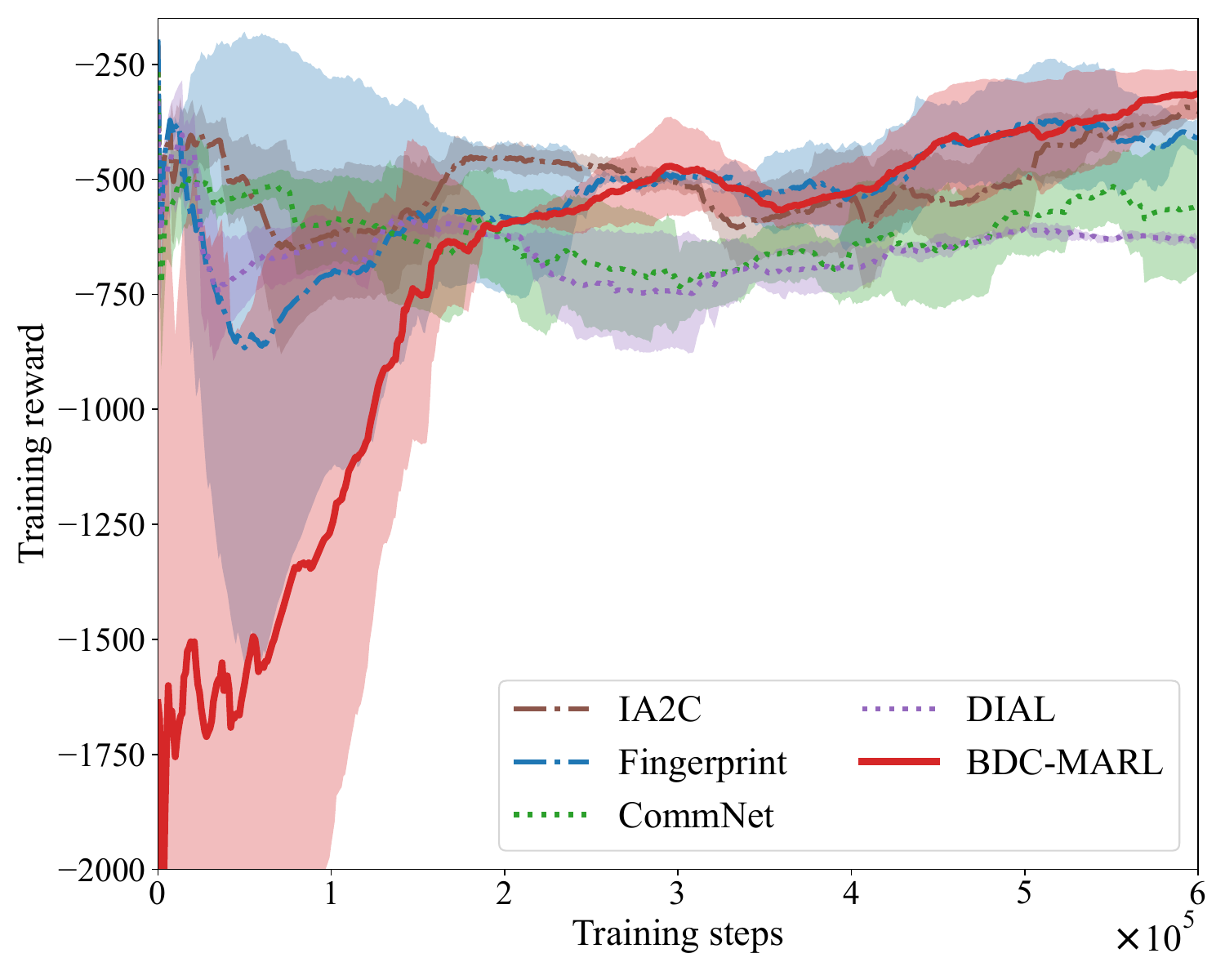}}
\caption{The comparison with different methods}
\label{fig2: marl_com}
\vspace{-10pt}
\end{figure}
A comparative learning progress through training curves for the various MARL algorithms are described as shown in Figure \ref{fig2: marl_com} and Table \ref{table1:reward comparison}. 
The proposed BDC-MARL outperforms other algorithms, starting with a higher initial reward and steadily increasing performance. Despite early fluctuations, it maintains a higher reward throughout training, indicating extensive strategy exploration before stabilizing. This thorough exploration can lead to more optimal policies. 
After the training phase, the evaluation phase for each algorithm included 50 distinct trials, each initiated with different initial conditions to ensure a thorough and varied testing environment, as shown in Table \ref{table2:comm comparison}. It achieved significantly higher rewards without compromising power consumption efficiency. Although it shows slightly higher standard deviations in some metrics, this suggests an adaptive strategy. All algorithms effectively prioritized safety, with consistently low or zero collisions.

\begin{table*}[!ht]
\centering
\caption{Reward comparison with different algorithms}
\label{table1:reward comparison}
\begin{tabular}{c|cccc|cccc|cc}
\hline \hline
Algorithms                           & \multicolumn{2}{c}{IA2C}    & \multicolumn{2}{c|}{Fingerprint} & \multicolumn{2}{c}{Commnet} & \multicolumn{2}{c|}{DIAL}    & \multicolumn{2}{l}{BDC-MARL} \\ \hline
Reward value                         & \multicolumn{2}{c}{-351.07} & \multicolumn{2}{c|}{-409.12}     & \multicolumn{2}{c}{-559.45} & \multicolumn{2}{c|}{-636.19} & \multicolumn{2}{c}{\textbf{-314.99}}    \\  \hline \hline
\end{tabular}
\end{table*}

\begin{table*}[!ht]
\centering
\caption{Test performance comparison with different algorithms}
\label{table2:comm comparison}
\begin{tabular}{c|cccc|cccc|cc}
\hline \hline
Algorithms                           & \multicolumn{2}{c}{IA2C}    & \multicolumn{2}{c|}{Fingerprint} & \multicolumn{2}{c}{Commnet} & \multicolumn{2}{c|}{DIAL}    & \multicolumn{2}{l}{BDC-MARL} \\ 
Statistical value                              & Avgerage        & Std       & Avgerage          & Std          & Avgerage       & Std        & Avgerage        & Std        & Avgerage         & Std         \\ \hline
IVS (m)                              & \textbf{20.75}           & 0.81      & 21.68             & 0.31         & 22.38          & 0.03       & 22.08           & 0.2        & 20.76            & 0.78        \\
Velocity (m/s)                       & 15.28           & 0.47      & 15.11             & 0.26         & \textbf{15.01}          & 0.03       & 15.06           & 0.09       & 15.26            & 0.69        \\
Acceeration (m/s\textasciicircum{}2) & 0.11            & 0.18      & 0.08              & 0.13         & \textbf{0.01}           & 0.02       & 0.04            & 0.05       & 0.15             & 0.21        \\
Platoon power consumption (kW)               & 17.97           & 0.31      & 18.28             & 0.42         & 18.94          & 0.04       & 18.89           & 0.15       & \textbf{17.85}            & 0.46        \\
Collision number         & 0           & 0      & 0             & 0         & 0          & 0       & 0           & 0       & 0            & 0       \\            
% & \multicolumn{2}{c}{0}       & \multicolumn{2}{c|}{0}           & \multicolumn{2}{c}{0}       & \multicolumn{2}{c|}{0}       & \multicolumn{2}{c}{0}          \\ 
\hline \hline
\end{tabular}
\end{table*}

\begin{figure*}
    \centering
    \subcaptionbox{IA2C method}[1.0\textwidth]{\includegraphics[width=1.0\textwidth]{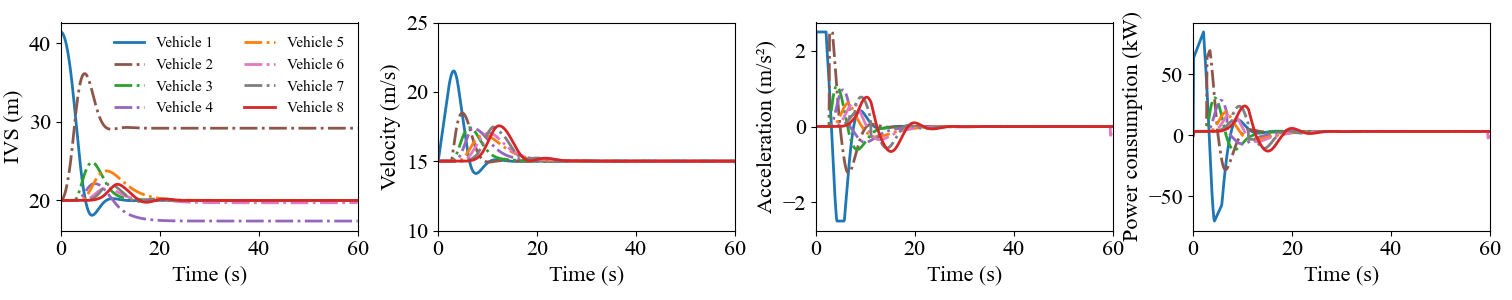}}
    % \subcaptionbox{IA2C method}[1.0\textwidth]{\includegraphics[width=1.0\textwidth]{Fig/ia2c_10.png}}
    % \subcaptionbox{Fingerprint method}[1.0\textwidth]{\includegraphics[width=1.0\textwidth]{Fig/fp_150.png}}
    \subcaptionbox{CommNet method}[1.0\textwidth]
    {\includegraphics[width=1.0\textwidth]{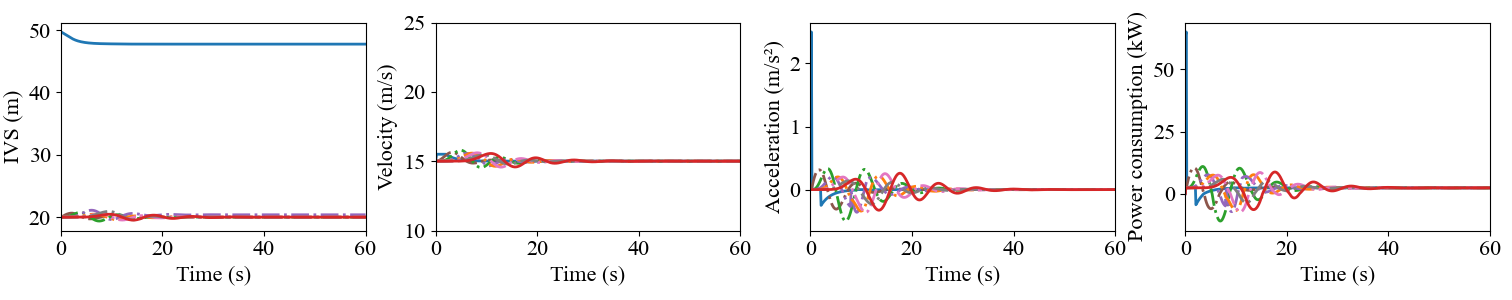}}
    % {\includegraphics[width=1.0\textwidth]{Fig/commnet_10.png}}
    \subcaptionbox{BDC-MARL method}[1.0\textwidth]
    {\includegraphics[width=1.0\textwidth]{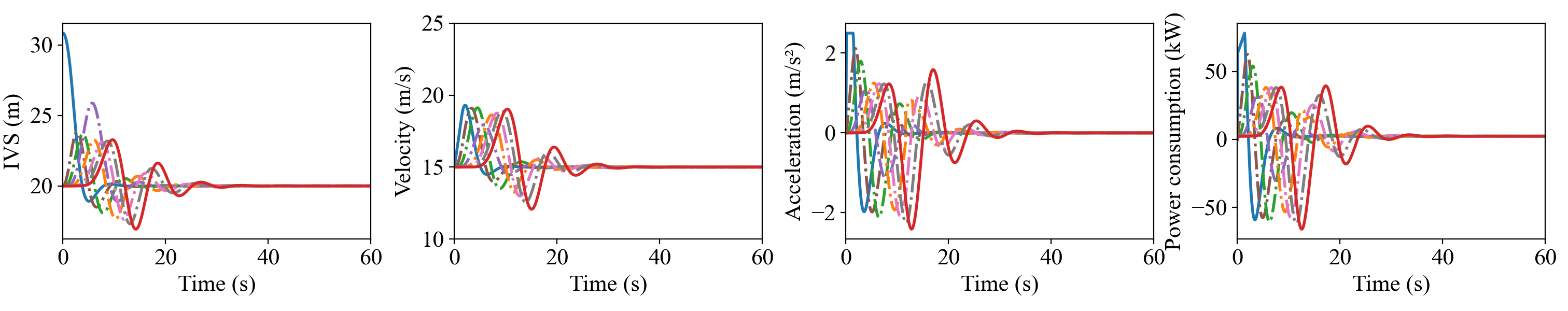}}
    % {\includegraphics[width=1.0\textwidth]{Fig/qcon_3.png}}
    \caption{The performance comparison of all vehicles with typical MARL algorithms}
    \label{fig3:marl_com_test}
\vspace{-15pt}
\end{figure*}
Figure \ref{fig3:marl_com_test} compares the performance of 8 vehicles using different MARL algorithms, including I2AC, CommNet, and BDC-MARL, based on IVS, velocity, acceleration, and power consumption over time. 
Divergences in some vehicles with I2AC (e.g., Vehicle 2 and 4) and CommNet (e.g., Vehicle 1) suggest challenges in information sharing and coordination. Effective communication is crucial for synchronizing actions and maintaining uniform IVS and velocity in MARL systems like CACC. The I2AC method’s divergences may indicate lagging communication channels, while CommNet’s velocity fluctuations suggest inefficient use of communicated information. These issues imply potential communication latency or protocols preventing agents from aligning actions effectively in dynamic tasks. Conversely, BDC-MARL’s convergence towards the desired IVS and velocity with less variability indicates a more efficient communication strategy, enabling better coordination and faster adaptation to achieve goals.
\vspace{-10pt}
\subsection{Communication Efficiency of MARL on CACC}
\begin{figure}
\centerline{\includegraphics[width=0.8\columnwidth]{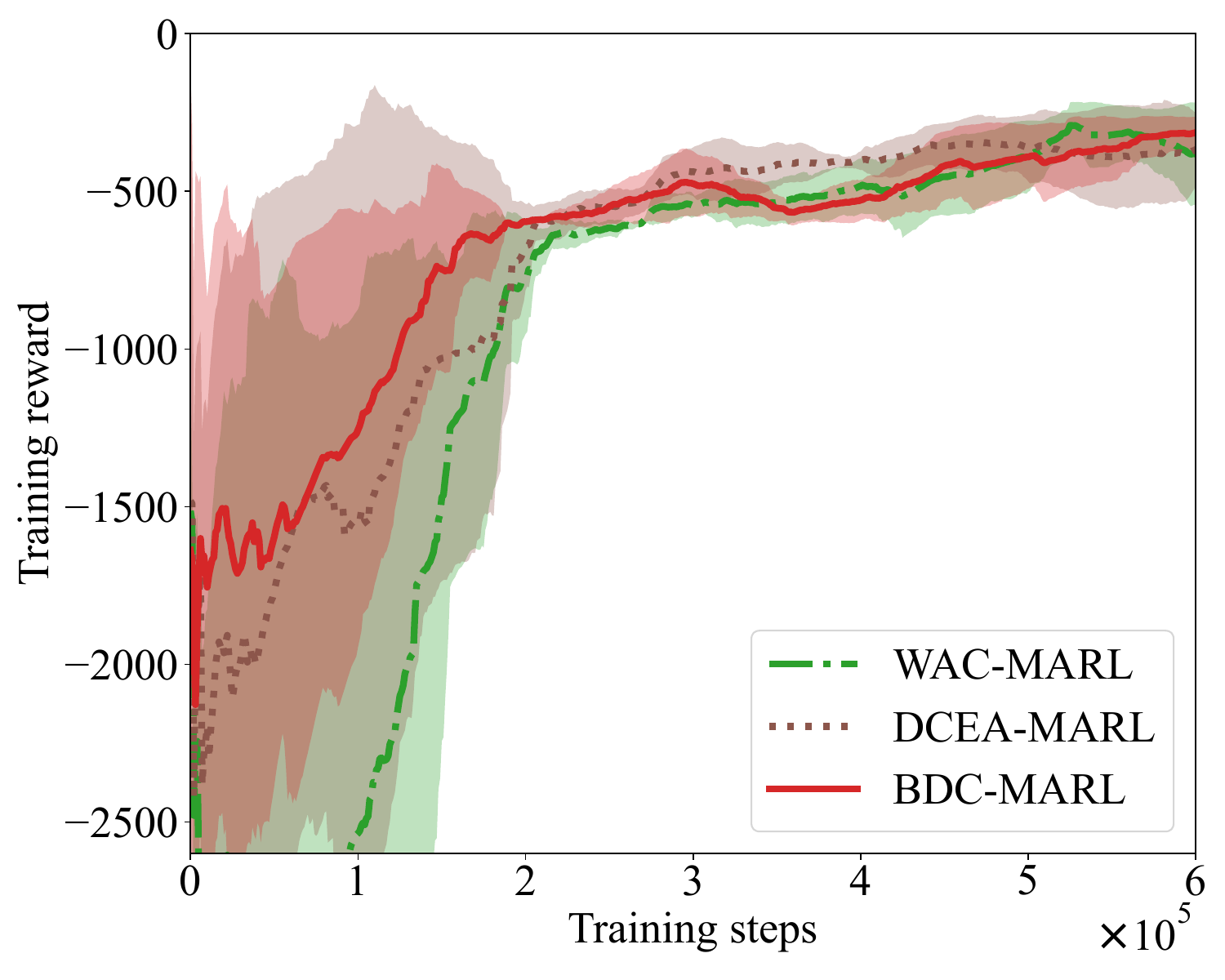}}
\caption{The comparison with three different sharing methods}
\label{fig4:sharing method}
\vspace{-10pt}
\end{figure}

Using different information-sharing methods in CACC systems is crucial because each method offers unique benefits in various traffic scenarios and communication environments. For example, a broadcast method ensures all vehicles have the same information, promoting consistent behavior, enhancing safety, and preventing collisions. This is achieved with weight averaging consensus (WAC)-MARL, where each agent’s weights are averaged with its neighbors. Alternatively, directed information sharing, which communicates with neighbors, efficiently relays information in both directions within a platoon. This approach uses differential consensus-enhanced adjustment (DCEA)-MARL, adjusting each agent’s weights by adding a weighted difference between the neighbors’ weights and its own. The term 
% \begin{equation}
$w_i^{\prime}=w_i+\epsilon \sum_{j \in \text {neighbors}(i)}\left(w_{j+1}-w_{i+1}\right)$
% \end{equation}
captures the essence of this update mechanism, where an epsilon factor $\epsilon$ is the scale factor. The selective sharing, such as the proposed method BDC-MARL, is more appropriate when the bandwidth is limited or when too much information could overwhelm the processing capabilities of individual agents, leading to increased reaction times or errors. Hence, Figure \ref{fig4:sharing method} presents a comparison of learning curves, highlighting the superior performance of the BDC-MARL method against the other two. 

\begin{figure*}
    \centering
    \subcaptionbox{WAC-MARL method}[1.0\textwidth]
    {\includegraphics[width=1.0\textwidth]{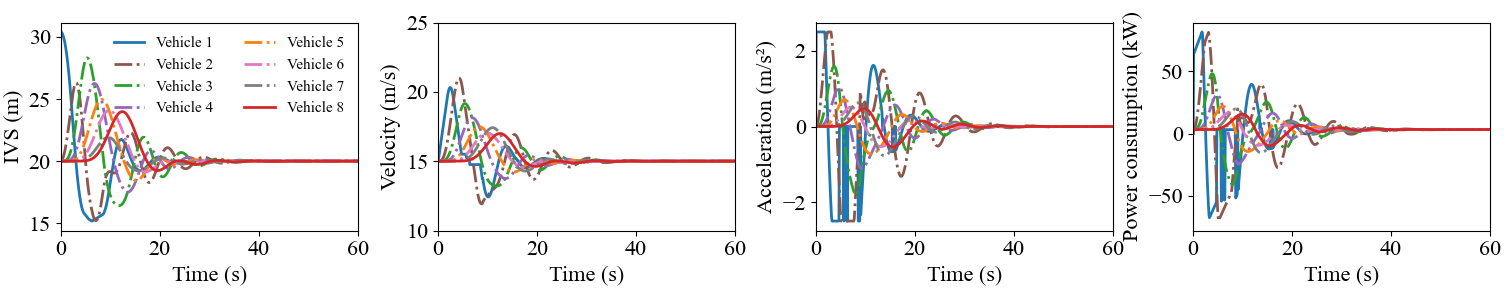}}
    % {\includegraphics[width=1.0\textwidth]{Fig/cu_100_4.png}}
    \subcaptionbox{DCEA-MARL method}[1.0\textwidth]
    {\includegraphics[width=1.0\textwidth]{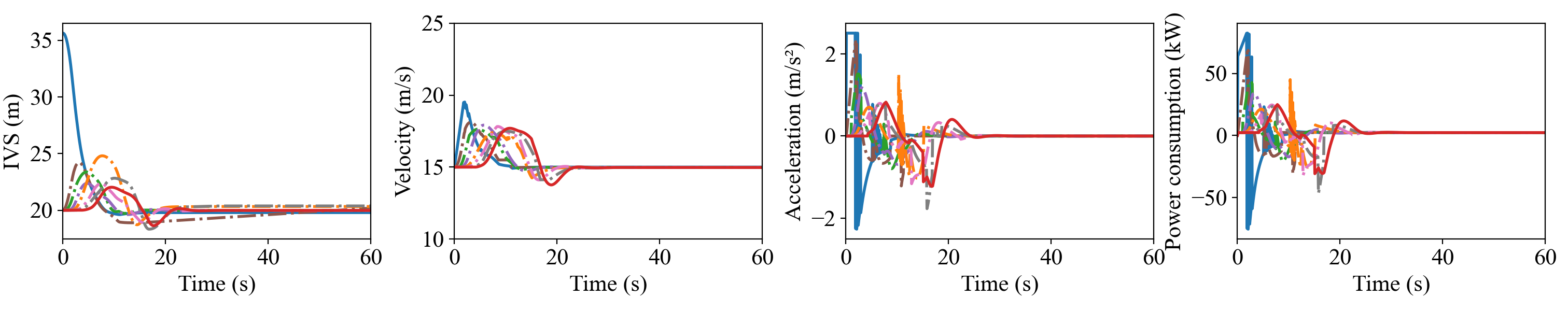}}
    % {\includegraphics[width=1.0\textwidth]{Fig/qcon_2_2.png}}
    \caption{The performance process of different sharing methods}
    \label{fig5:all performence}
\vspace{-10pt}
\end{figure*}

While BDC-MARL exhibits some mid-training variability, it ultimately achieves the highest rewards, making it the most effective method for maximizing training rewards. DCEA-MARL shows good initial learning velocity and stability, ideal for scenarios where early performance is critical. WAC-MARL demonstrates steady improvement and stability, suitable for environments requiring consistent performance gains over time.
After the training phase, each algorithm was evaluated on 50 seeds under varied starting conditions. The performance process with all vehicles is illustrated in Figure \ref{fig5:all performence}. Comparing Figure \ref{fig3:marl_com_test} (c) with Figure \ref{fig5:all performence}, all three methods eventually converge towards a pre-designed IVS and velocity. The WAC method shows initial over-corrections in acceleration and power consumption before stabilizing. The DCEA method reaches stability quicker than the WAC method, despite initial inconsistencies in acceleration and velocity. The BDC-MARL approach quickly converges in terms of IVS and maintains a more stable velocity and acceleration pattern, due to its efficient use of limited bandwidth and information processing capabilities, preventing data overflow and response lag.

\begin{figure}[h]
\centerline{\includegraphics[width=1.0\columnwidth]{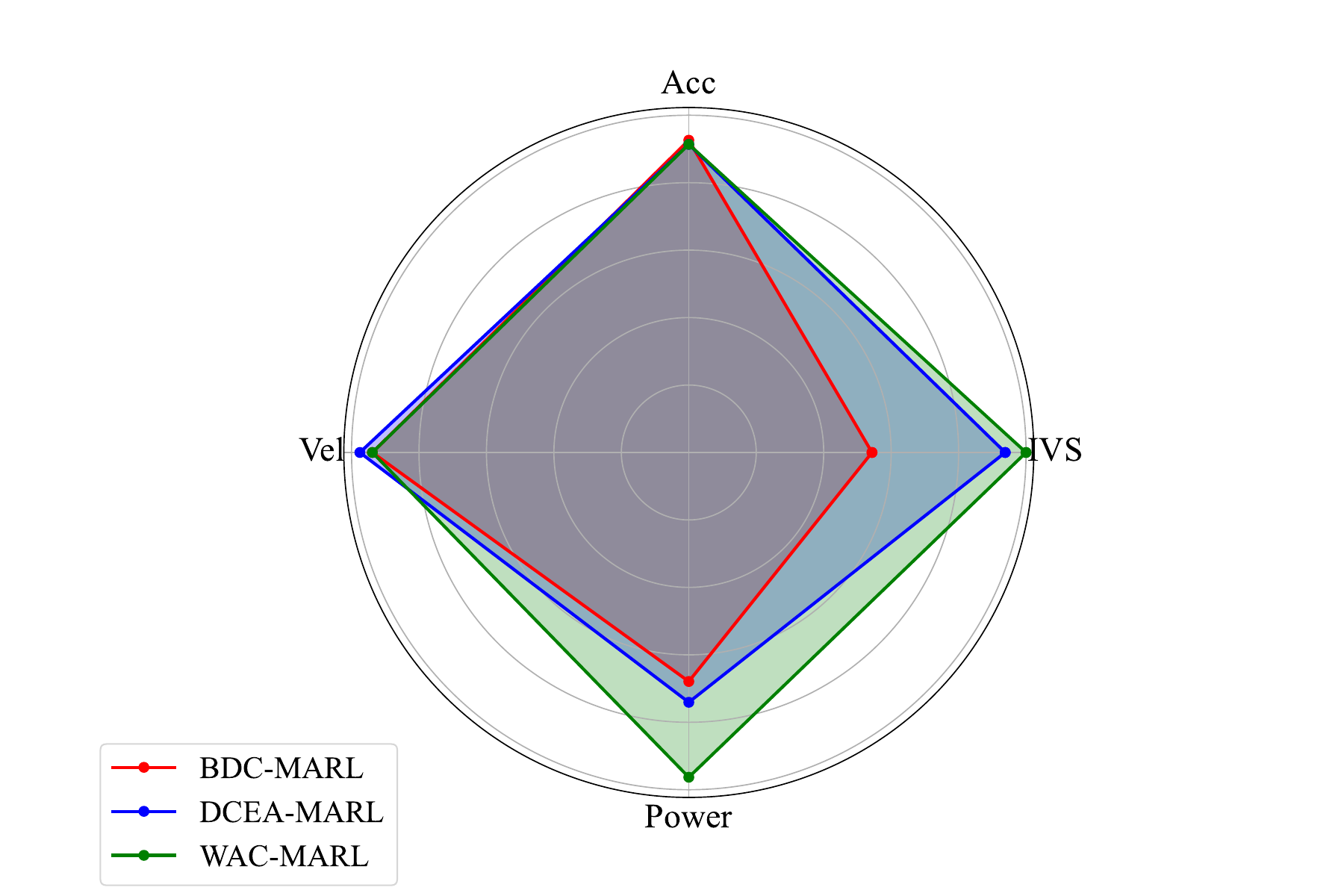}}
\caption{The statistical comparison of different communicative methods}
\label{sharing_method}
\vspace{-10pt}
\end{figure}

The radar chart in Figure \ref{sharing_method} evaluates the performance of three methods using four metrics. This chart effectively visualizes comprehensive performance through the aggregate area of each method’s plot. A smaller area within the radar chart signifies a more robust overall performance, indicating a method's efficiency and effectiveness across the evaluated dimensions. Therefore, the balance across all four metrics for BDC-MARL can imply a more robust overall system performance.
For IVS and velocity, all three methods show comparable performance, with their lines relatively close to the outer edge. However, in acceleration and power consumption, the BDC-MARL method is noticeably closer to the edge, indicating superior performance in these areas. 

In summary, while the WAC and DCEA methods initially face oscillations, they eventually achieve consensus, promoting collective efficiency. However, the BDC-MARL method stands out for its quick convergence and stability, demonstrating its effectiveness in environments with limited communication resources.

\vspace{-10pt}
\subsection{Impact of Platooning Size on the Scalability of MARL on CACC}
In CACC, the overall performance including energy efficiency, safety, and the effectiveness of communication will be affected by different vehicle sizes, since there is the propagation phenomenon in the response and communication signals.

\begin{figure}[h!]
\centerline{\includegraphics[width=0.8\columnwidth]{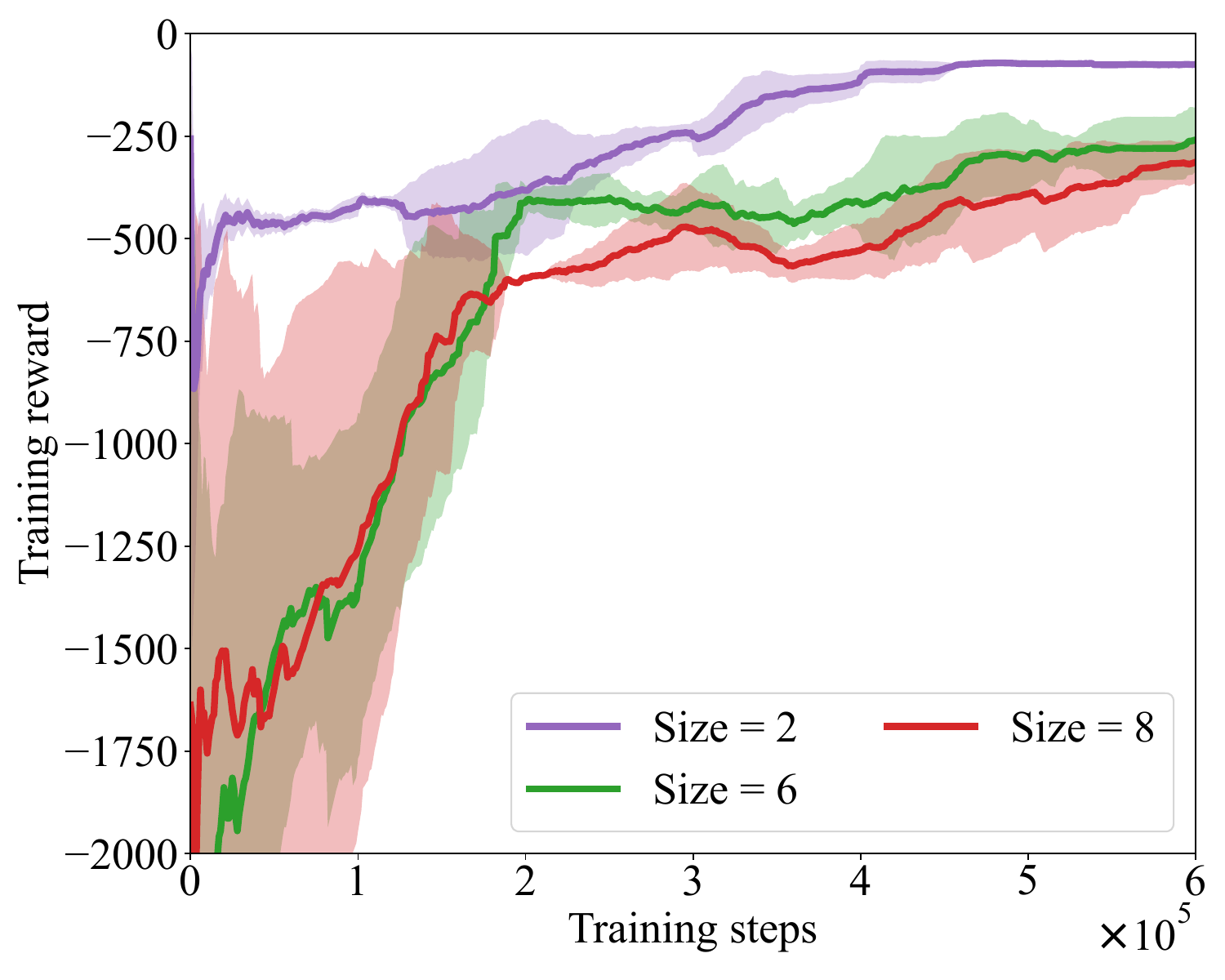}}
\caption{The comparison with different sizes}
\label{fig7:marl_size}
\vspace{-10pt}
\end{figure}

Figure \ref{fig7:marl_size} shows the training curves with different sizes, ranging from a minimal arrangement (i.e., 2 AVs) to more populated configurations (i.e., 6 and 8 AVs), thereby providing a comprehensive overview of the scalability and adaptability in a variety of urban traffic scenarios, since larger vehicle sizes experience more complex effects, such as increased air resistance or more intricate behavior in a platoon. In Figure \ref{fig7:marl_size}, we can find that Size 8 needs more sophisticated models to handle these effects effectively, which could explain a slower but steady increase in training process. In contrast, Size 2 experiences less interaction and therefore might find an optimal policy faster.

\begin{figure}[h!]
\centerline{\includegraphics[width=1.0\columnwidth]{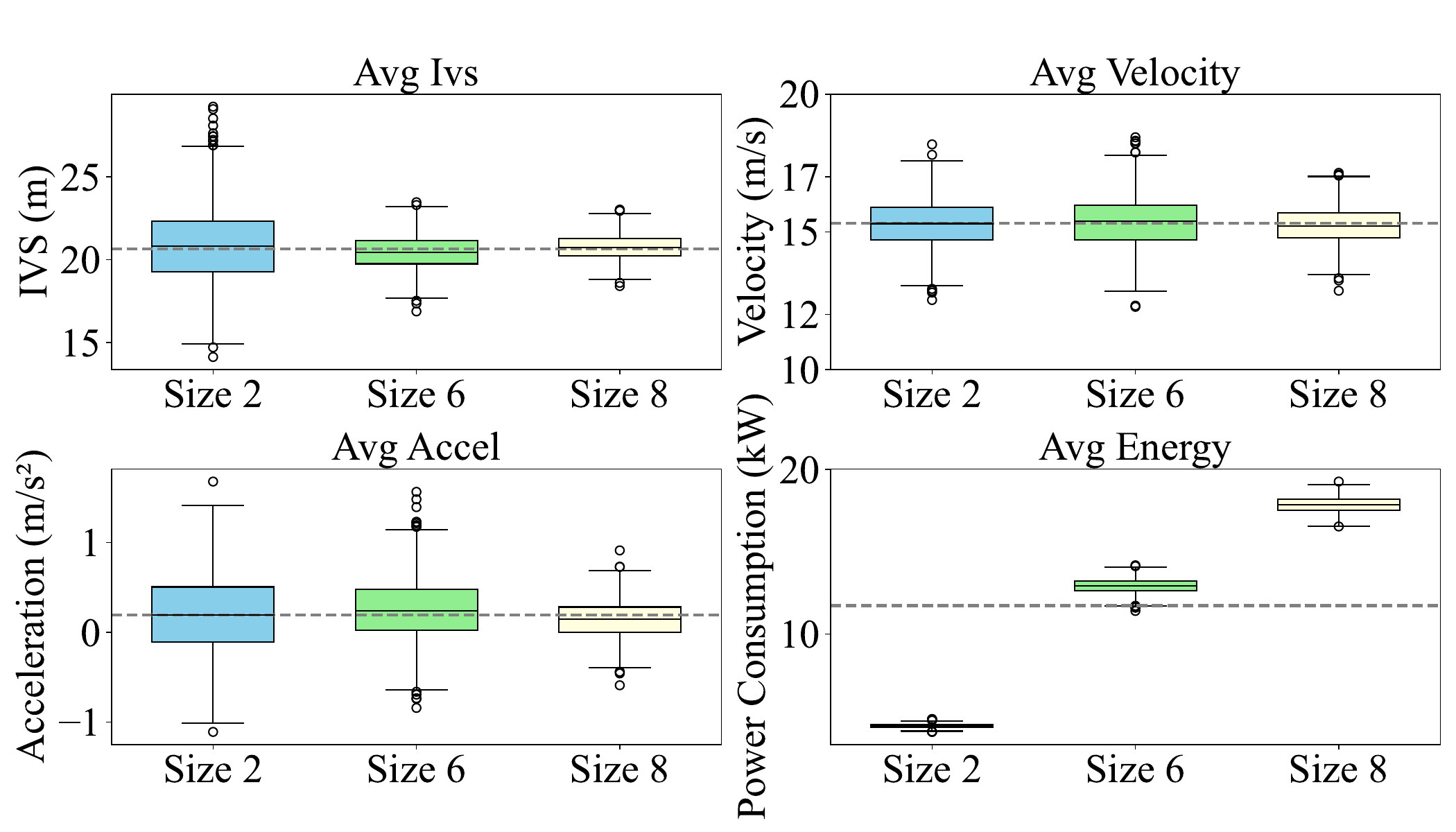}}
\caption{The comparison with different size}
\label{fig8:size_test}
\vspace{-10pt}
\end{figure}

The overall test performance under 50 different trials is displayed in Figure \ref{fig8:size_test}. 
The comprehensive analysis of varying platoon sizes suggests that while larger platoons achieve fuel savings due to decreased aerodynamic drag, this does not result in a proportional increase in power consumption. Interestingly, a platoon size of 6 emerges as the most advantageous, exhibiting consistent IVS, moderate velocities, and the lowest variation in power consumption, indicative of an efficient balance between fuel efficiency and operational stability. Although larger platoons benefit from aerodynamic efficiency, they also confront the challenge of maintaining stable and safe following distances amidst dynamic traffic conditions, which is less of a concern with smaller platoons. Therefore, a platoon size of 6 is potentially optimal for achieving a harmonious balance between power efficiency, vehicle spacing stability, and the complexities inherent in managing larger numbers of closely following vehicles.
More importantly, it also indicates that the presented algorithms effectively optimize platoon dynamics, achieving fuel economy improvements without a corresponding escalation in power consumption across varying platoon sizes.

\subsection{Co-optimization Performance Validation with Real-World Data}
In this part, the different targets (the target IVS and velocity) on the performance of the presented algorithm has been conducted from the real-data scenarios extracted from OpenACC.
% , as shown in Table \ref{openacc scenarios}.
The examination is primarily focused on analyzing three different IVS settings with the target velocity 11 $m/s$ and the Size 4, ranging from IVS setting $\mathrm{S}$ $(12m)$, $\mathrm{M}$ $(16m)$, and $\mathrm{L}$ $(20m)$, labelled as Group 1, Group 2, and Group 3. 

% As shown in Figure \ref{appen6: cacc_comparison} of Appendix, the comparisons between the BDC-MARL strategy and OpenACC data across three groups are conducted, 
% BDC-MARL strategy maintains the target IVS with minor deviations, especially in the setting with the largest IVS (Group 3). Since larger IVS could introduce greater variances due to increased reaction time and distance, making the system slightly less accurate in maintaining a consistent gap.
% The test performance in the real-data scenario for different groups is relatively close in metrics velocity and acceleration, the BDC-MARL algorithm with less delay effectively balances acceleration and braking to match the target velocity despite different spacing settings.

Table \ref{table3:openacc results} presents average and standard deviation values for velocity, IVS, acceleration, and total energy for vehicles in the proposed BDC-MARL algorithm and real-world OpenACC data. Here, the average velocities are close to the target velocity of 11 m/s, indicating that all strategies and groups are performing well in terms of pre-designed setting. However, there are noticeable differences in the standard deviation of IVS, showing greater variability in the OpenACC. The total energy consumption is fairly consistent across groups,  since the analysis has been conducted over a short period, which might not be sufficient to capture the long-term benefits of BDC-MARL in terms of energy savings but the stability of the vehicle dynamics is improved.

\begin{table*}[h!]
\centering\caption{Extraction of real-world scenarios from OpenACC}
\label{table3:openacc results}
\begin{tabular}{ccccccc}
\hline \hline
Group              & Statistical Values       & Method  & Velocity  (11 m/s) & IVS (m) & Acceleration & Total energy for all vehicles  (kWh) \\ \hline
\multirow{4}{*}{1 (S)} & \multirow{2}{*}{Average} & OpenACC & 10.81           & 16.96   & 0.01         & 0.11                                                               \\
                   &                          & BDC-MARL  & 10.75           & \textbf{11.99}   & \textbf{0}            & 0.11                                                               \\
                   & \multirow{2}{*}{Std}     & OpenACC & 2.01            & 5.36    & 0.77         &     /                                                               \\
                   &                          & BDC-MARL  & \textbf{0.73}            & \textbf{1.37}    & \textbf{0.23}         &   /                                                                 \\ \hline
\multirow{4}{*}{2 (M)} & \multirow{2}{*}{Average} & OpenACC & 10.28          & 21.89   & 0.11         & 0.09                                                               \\
                   &                          & BDC-MARL  & 10.2            & \textbf{16.99}   & \textbf{0}            & 0.09                                                               \\
                   & \multirow{2}{*}{Std}     & OpenACC & \textbf{2.08}            & 3.68    & \textbf{0.68}         &   /                                                                 \\
                   &                          & BDC-MARL & 2.23            & \textbf{2.27}    & 0.71         &   /                                                                 \\ \hline
\multirow{4}{*}{3 (L)} & \multirow{2}{*}{Average} & OpenACC & 10.6            & 25.12   & 0.02         & 0.10                                                               \\
                   &                          & BDC-MARL  & 10.47           & \textbf{19.54}   & \textbf{0}            & 0.10                                                               \\
                   & \multirow{2}{*}{Std}     & OpenACC & 1.68            & 4.3     & \textbf{0.55}         &   /                                                                 \\
                   &                          & BDC-MARL  & \textbf{1.47}            & \textbf{2.15}    & \textbf{0.55}         &     /                                                               \\ \hline \hline
\end{tabular}
\end{table*}

\section{Conclusion}
The paper introduces a binary differential consensus (BDC)-MARL approach to enhance system robustness and efficiency. By minimizing dependency on potentially delayed or unreliable information exchanges between vehicles, this approach optimizes individual vehicle responses to real-time traffic conditions. The main conclusions are summarized as follows::
% \begin{enumerate}
% 可行的
% 缺点为什么
% openacc 

% 主句突出 有收敛优化 可行
% 可行之后 省油多少
% 第三是什么方法，有什么缺点，为什么能省油
% \item

(1) Compared with state-of-the-art algorithms, the BDC-MARL achieves the largest improvement in energy savings, up to 5.8\%, with an average velocity of 15.26 m/s and an IVS of 20.76 m. In CACC scenarios, delayed messages in real-time CACC tasks hinder performance, as observed with I2AC and CommNet. 
In contrast, BDC-MARL demonstrates efficient communication that effectively mitigates this issue. 
    % \item Despite initial fluctuations, WAC and DCEA methods achieve efficiency over time, but the quick stabilization of BDC method highlights its advantage in communication-limited CACC scenario.
% \item 

(2) Integrating QSGD into our communication protocol achieves stable performance and comparable energy savings across three information-sharing methods. Among them, the BDC-MARL approach rapidly converges for IVS and maintains a more stable velocity and acceleration pattern, effectively using limited bandwidth and processing capabilities to prevent data overflow and delays.

% Three information-sharing methods achieve stable performance and comparable energy savings, demonstrating the effective integration of QSGD into our communication protocol. However, the quickest stabilization of the BDC method highlights its advantage in communication-limited CACC scenario.

% \item 

(3) A 6-vehicle platoon is optimal for balancing efficiency and stability, demonstrating the scalability in optimizing platoon dynamics for energy efficiency. Larger platoons save fuel through reduced aerodynamic drag without increasing power use proportionally. 
    % Larger platoons save fuel through reduced aerodynamics drag without raising power use proportionally, with a 6-vehicle platoon optimal for balancing efficiency and stability, reflecting the algorithms' success at optimizing platoon dynamics for fuel economy without increased power costs.
% \item 

(4)While average velocities align with the target in both BDC-MARL and OpenACC data, OpenACC exhibits greater variability in IVS due to less efficient coordination between vehicles. Despite short-term consistency in energy consumption, BDC-MARL enhances vehicle dynamics stability by leveraging robust communication protocols and optimized response strategies.
% \end{enumerate}

Future work will focus on extending the analysis to longer periods to better assess the long-term energy savings potential of BDC-MARL, refining communication protocols to reduce variability in IVS, and enhance real-time coordination.

\section{Acknowledgement}
The work is supported in part by the Fundamental Research Funds for the Central Universities (22120240223), by the EPSRC (EP/J00930X/1), and by Innovate UK (102253).
\bibliographystyle{IEEEtran}
\bibliography{Refs}

% \clearpage 
% \newpage

% \input{Appendix}
% \begin{thebibliography}{00}

% \end{thebibliography}

\end{document}